\documentclass[floatfix,twocolumn,aps,prl,reprint,nofootnoteinbib]{revtex4-1}
\usepackage{graphicx}
\usepackage{amsmath}
\usepackage{amssymb}
\usepackage{hyperref}
\usepackage{bm}

\newcommand{\inleva}[1]{\langle#1\rangle}

\newcommand{\abs}[1]{\ensuremath{\left| #1 \right|}}

\newcommand{\Tr}{\ensuremath{\mathrm{Tr}}}

\newcommand{\fivevec}[5]{\ensuremath{\left(
  \begin{array}{c}#1\\#2\\#3\\#4\\#5\end{array}\right)}}
\newcommand{\threemat}[9]{\left(\begin{array}{ccc}#1&#2&#3\\#4&#5&#6\\#7&#8&#9\end{array}\right)}

\newcommand{\absF}{|\langle\mathbf{\hat{F}}\rangle|}
\newcommand{\expF}{\langle\mathbf{\hat{F}}\rangle}
\newcommand{\SO}{\mathrm{SO}}
\newcommand{\SU}{\ensuremath{\mathrm{SU}}}
\newcommand{\U}{\ensuremath{\mathrm{U}}}
\newcommand{\D}{\ensuremath{\mathrm{D}}}
\newcommand{\M}{\ensuremath{\mathcal{M}}}
\newcommand{\rr}{\ensuremath{\mathbf{r}}}

\newcommand{\R}{\ensuremath{\mathcal{R}}}
\newcommand{\I}{\ensuremath{\mathcal{I}}}
\newcommand{\C}{\ensuremath{\mathcal{C}}}
\newcommand{\Ct}{\ensuremath{\tilde{\C}}}
\newcommand{\chiB}{\ensuremath{\chi_\mathrm{B}}}
\newcommand{\zBN}{\ensuremath{\zeta^\mathrm{BN}}}
\newcommand{\Dt}{\ensuremath{\tilde{\D}}}
\newcommand{\sx}{\ensuremath{\sigma_x}}
\newcommand{\sy}{\ensuremath{\sigma_y}}
\newcommand{\sz}{\ensuremath{\sigma_z}}

\newcommand{\rhat}{\ensuremath{\mathbf{\hat{r}}}}
\newcommand{\xhat}{\ensuremath{\mathbf{\hat{x}}}}
\newcommand{\yhat}{\ensuremath{\mathbf{\hat{y}}}}
\newcommand{\zhat}{\ensuremath{\mathbf{\hat{z}}}}
\newcommand{\nhat}{\ensuremath{\mathbf{\hat{n}}}}

\begin{document}

\author{Magnus O.\ Borgh}
\altaffiliation[Current address: ]{Faculty of Science, University of
  East Anglia, Norwich, NR4 7TJ, United Kingdom}
\affiliation{Mathematical Sciences, University of Southampton, SO17 1BJ,
    Southampton, United Kingdom}
\author{Janne Ruostekoski}
\affiliation{Mathematical Sciences, University of Southampton, SO17 1BJ,
    Southampton, United Kingdom}

\date{\today}

\title{Core Structure and Non-Abelian Reconnection of Defects in a Biaxial Nematic Spin-2 Bose-Einstein Condensate}

\begin{abstract}
We calculate the energetic structure of defect cores and propose
controlled methods to imprint a nontrivially entangled vortex pair
that undergoes non-Abelian vortex reconnection in a biaxial nematic spin-2
condensate. For a singular vortex, we find three superfluid cores in
addition to depletion of the condensate density.
These exhibit order parameter symmetries that are different
from the discrete symmetry of the biaxial nematic phase, forming an
interface between the defect and the bulk superfluid.  We provide a
detailed analysis of phase mixing in the resulting vortex cores and
find an instability dependent upon the orientation of the order
parameter.
We further show that the spin-2 condensate is a promising system for
observing spontaneous deformation of a point defect into an ``Alice
ring'' that has so far avoided experimental detection.
\end{abstract}

\maketitle

Topological defects and textures are
ubiquitous across physical systems that seemingly have little in
common~\cite{pismen}, from
liquid crystals~\cite{kleman} and superfluids~\cite{volovik} to cosmic
strings~\cite{vilenkin-shellard}. They arise generically from
symmetries of a ground state that is described by an
order parameter~\cite{mermin_rmp_1979}---a function parametrizing the
set of physically distinguishable, energetically degenerate states.
In the simple example of a scalar superfluid, the order parameter is the
phase of the macroscopic wave function.  More generally it may be a vector or
tensor that is symmetric under particular transformations.
In a uniaxial nematic (UN), the order parameter is cylindrically symmetric
around a locally defined axis.
It also exhibits a twofold
discrete symmetry under reversal of the cylinder axis, which leads to
half-quantum vortices (HQVs) in atomic spinor Bose-Einstein condensates
(BECs)~\cite{leonhardt_jetplett_2000,zhou_ijmpb_2003,lovegrove_pra_2012,seo_prl_2015} and
superfluid liquid $^3$He~\cite{vollhardt-wolfle}, and to
$\pi$ disclinations in liquid crystals~\cite{kleman,pismen}. In a
biaxial nematic (BN), also the cylindrical symmetry is broken into the
fully discrete symmetry of a rectangular brick, with dramatic
consequences: the BN is the simplest order parameter
that supports non-Abelian vortices that
do not commute~\cite{poenaru_jphys_1977,mermin_rmp_1979}. As a result,
colliding non-Abelian vortices cannot reconnect
without leaving traces of the process, but must instead form a
connecting rung vortex.
Noncommuting defects appear as cosmic strings in theories of the
early universe~\cite{vilenkin-shellard},
and have been predicted in
BN liquid crystals~\cite{poenaru_jphys_1977}.

Despite long-standing experimental efforts,
BN phases in liquid crystals have experimentally proved more elusive
than originally anticipated~\cite{luckhurst-sluckin}.
In atomic systems, the topological classification and dynamics of
non-Abelian vortices have theoretically been studied
in the cyclic phase of
spin-2~\cite{semenoff_prl_2007,huhtamaki_pra_2009,kobayashi_prl_2009,mawson_pra_2015}
and in spin-3 BECs~\cite{barnett_pra_2007},
though it
remains uncertain whether any alkali-metal atoms
exhibit the corresponding ground states.
Consequently, any physical system where non-Abelian defects
may be reliably studied is still lacking.

Spin-2 BECs exhibit---in addition to the ferromagnetic (FM) and cyclic
phases---both UN and BN
phases~\cite{turner_prl_2007,song_prl_2007,kawaguchi_physrep_2012},
which, however, are degenerate at the mean-field level.
Beyond mean-field
theory, the degeneracy may be lifted by quantum
fluctuations through interaction-dependent ``order-by-disorder''
processes~\cite{song_prl_2007,turner_prl_2007}.

Here, we instead break the degeneracy already at the mean-field level with a
Zeeman shift and compute the energetically (meta)stable defect
structures as well as propose a scheme for experimentally preparing
vortices exhibiting non-Abelian reconnection. This shows how
BN spin-2 BECs can provide an experimentally simple path to study
structure and dynamics of non-Abelian defects.
For an energetically stable singular vortex in a rotating system,
we fully characterize the
appearance of three superfluid core structures, in addition to
the empty core.  As well as one FM and one cyclic order-parameter
core, a third superfluid core appears that
exhibits a twofold symmetry, with cyclic and BN phases appearing
offset inside the core. We provide a parametrization of the vortex
wave function that captures the symmetry breaking and mixing of phases
in the core.
The structure of the superfluid cores results from a
combination of energetics and topology as an interface forms between
the BN phase of the bulk superfluid and the different symmetry of the
core order parameter.
Considering the cyclic core, we find that the discrete point group
symmetry not only manifests itself in an anisotropic core shape, but
also leads to an orientation-dependent instability, when the vortex
line is sufficiently tilted with respect to the orientation of the
order parameter.

The study of non-Abelian vortex dynamics requires preparation of a
vortex pair with noncommuting topological charges.
The order-parameter transformations corresponding to each vortex
combine nontrivially, leading to a complicated wave function that
is not amenable to conventional imprinting techniques.
We propose a two-step protocol to imprint the vortex pair,
based on rotating the effective magnetic
field between the preparation of each vortex. This allows the imprinting
to make use of simple representations of each vortex in a changing
spinor basis.
We numerically demonstrate the non-Abelian vortex reconnection.
The proposed method could be generalized to controlled creation
of systems of noncommuting vortices in atomic superfluids.

We also show how a BN spin-2 BEC is a promising system to observe spontaneous
deformation of a point defect into a HQV ring, called an ``Alice ring''~\cite{ruostekoski_prl_2003},
whose detection has been beyond experimental resolution in spin-1 BECs despite considerable efforts~\cite{ray_nature_2014,tiurev_pra_2016}.
Alice rings are originally known from high-energy physics~\cite{schwarz_npb_1982,alford_npb_1991}
and represent a direct consequence of the topological ``hairy-ball theorem.''

Experimental interest in topological defects and textures in spinor BECs
is currently accelerating.
Recent efforts in spin-1 BECs have led to the \emph{in situ}
observation of a singly
quantized vortex splitting into a pair of HQVs~\cite{seo_prl_2015},
confirming theoretical
prediction~\cite{lovegrove_pra_2012}, and to controlled preparation of
coreless-vortex
textures~\cite{leanhardt_prl_2003,leslie_prl_2009,choi_prl_2012}, the
analogs of Dirac~\cite{ray_nature_2014} and
't~Hooft--Polyakov~\cite{ray_science_2015} monopoles, and particlelike
solitons~\cite{hall_nphys_2016}.
Our results for spin-2 reveal a defect-structure phenomenology
considerably richer than that in the spin-1
BECs~\cite{yip_prl_1999,mizushima_pra_2002,ji_prl_2008,lovegrove_pra_2012,lovegrove_pra_2016}.
Preparation of spin-2 BECs has been achieved using
$^{87}$Rb~\cite{schmaljohann_prl_2004,chang_prl_2004,kuwamoto_pra_2004,leslie_prl_2009},
which, like $^{23}$Na, is predicted to exhibit
the nematic phases~\cite{kawaguchi_physrep_2012,widera_njp_2006}.
The BN phase can then be realized by controlling the quadratic level shift,
e.g., by microwave dressing~\cite{gerbier_pra_2006} or laser
fields~\cite{santos_pra_2007}.
Both $^{87}$Rb and $^{23}$Na are commonly used in spinor-BEC
experiments, making the BN
phase the most likely candidate for realization of a ground-state
manifold supporting non-Abelian defects.

We write the mean-field-theoretical spin-2 condensate wave function in
terms of the density $n(\rr)$ and a normalized spinor $\zeta(\rr)$ as
$\Psi(\rr) =
\sqrt{n(\rr)}(\zeta_{+2}(\rr),\zeta_{+1}(\rr),\zeta_{0}(\rr),\zeta_{-1}(\rr),\zeta_{-2}(\rr))^T$.
The Hamiltonian density then reads~\cite{kawaguchi_physrep_2012}
\begin{equation}
  \label{eq:hamiltonian}
    \mathcal{H} =
    h_0
    + \frac{c_0}{2}n^2
    + \frac{c_2}{2}n^2\absF^2
    + \frac{c_4}{2}n^2\abs{A_{20}}^2
    + \mathcal{H}_Z(\expF),
\end{equation}
where
$h_0=(\hbar^2/2m)\abs{\nabla\Psi}^2 + (m\omega^2r^2/2)n$,
for atomic mass $m$ and an isotropic trap with frequency $\omega$.
The spin operator $\mathbf{\hat{F}}$ is given by a vector of spin-2 Pauli
matrices.  In addition to the spin-independent and $\absF^2$-dependent
interaction energies, a third interaction term
arises proportional to
$|A_{20}(\rr)|^2 = \frac{1}{5}\left|2\zeta_{+2}\zeta_{-2} - 2\zeta_{+1}\zeta_{-1}
  + \zeta_0^2\right|^2$, where $A_{20}$ is the amplitude of
spin-singlet pair formation~\cite{ueda_pra_2002}. The interaction
strengths are given by the $s$-wave scattering lengths $a_f$
in the spin-$f$ channels of colliding spin-2 atoms as
$c_0 = 4\pi\hbar^2(3a_4+4a_2)/7m$, $c_2 = 4\pi\hbar^2(a_4-a_2)/7m$, and
$c_4 = 4\pi\hbar^2(3a_4-10a_2+7a_0)/7m$.
Finally, $\mathcal{H}_Z(\expF) = pn\inleva{\hat{F}_z} +
qn\inleva{\hat{F}_z^2}$ represents linear and quadratic Zeeman
shifts of strengths $p$ and $q$, respectively.

When the Zeeman shifts are small, the spin-2 BEC exhibits three
distinct, interaction-dependent ground-state
phases~\cite{ciobanu_pra_2000,ueda_pra_2002,kawaguchi_physrep_2012}. These
may be characterized in terms of $\absF$ and $|A_{20}|$:
a spin-2 FM phase with $\absF=2$ and $|A_{20}|=0$, a cyclic
phase with $\absF=|A_{20}|=0$, and the polar phase with $\absF=0$ and
$|A_{20}|^2=1/5$.
A polar spinor can be written as both
$\zeta^\mathrm{UN} = (0,0,1,0,0)^T$ and
$\zBN = (1/\sqrt{2},0,0,0,1/\sqrt{2})^T$, representing the UN and BN
phases, respectively.
These are energetically degenerate for
$p=q=0$~\cite{turner_prl_2007,song_prl_2007}.
They may be distinguished by the amplitude of
spin-singlet trio formation~\cite{ueda_pra_2002}
$A_{30} = \left(3\sqrt{6}/2\right)
\left(\zeta_{+1}^2\zeta_{-2} + \zeta_{-1}^2\zeta_{+2}\right)
+ \zeta_0\left(\zeta_0^2 - 3\zeta_{+1}\zeta_{-1} -
  6\zeta_{+2}\zeta_{-2}\right)$, taking values
$|A_{30}|^2=0$, $1$, and $2$ in BN, UN, and cyclic phases, respectively.

\begin{figure}[tb]
  \centering
  \includegraphics[width=\columnwidth]{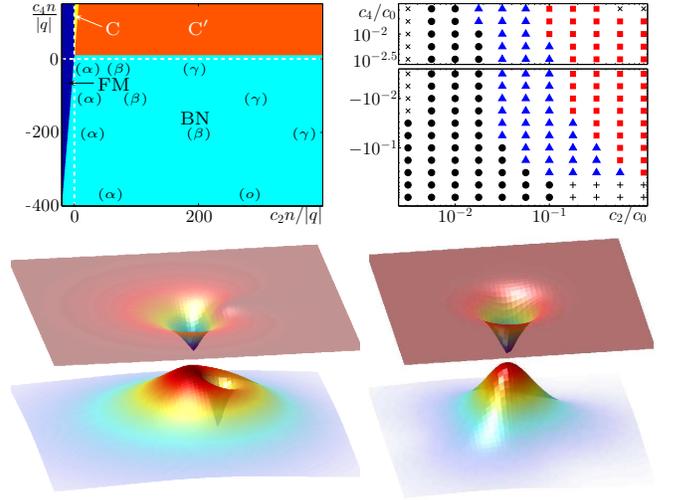}
  \caption{Top: Ground-state phase diagram for
  $q<0$ with cores of $(1/2,\sigma)$ vortex indicated (left) and
  detailed core phase diagram (right): [($o$)/$+$] empty,
  [$(\alpha)$/$\bullet$] FM, [$(\beta)$/$\blacktriangle$] cyclic,
  [$(\gamma)$/$\blacksquare$] coexistence of cyclic and
  phase-mixing cores; ($\times$) single vortex unstable. Bottom: $|A_{20}|$
  (top surface) and $|A_{30}|$ (bottom surface) in the phase-mixing
  (left) and cyclic (right) cores.
}
\label{fig:cores}
\end{figure}

With a nonzero quadratic Zeeman shift along the $z$ direction,
$\zBN$ acquires an energy $\mathcal{H}_Z = 4qn$, while
the energy of $\zeta^\mathrm{UN}$ is unchanged, breaking the
degeneracy and energetically favoring the BN phase whenever $q<0$. By
controlling the quadratic level
shift~\cite{gerbier_pra_2006,santos_pra_2007}, the BN phase could be
realized by experimentally simple means.
The ground-state phase diagram is shown in
Fig.~\ref{fig:cores}.  The BN phase is the ground state
for $c_2 > c_4/20$ and $c_4 < 10|q|/n$, bordering the FM phase and two
phases that continuously become cyclic for
$|q|\to0^-$~\cite{kawaguchi_physrep_2012,supplemental}.

Axisymmetric vortices in the polar interaction regime were studied in
Ref.~\cite{pogosov_pra_2005}.
With the energetic degeneracy between UN and BN phases lifted, we now
separately consider their defects.
While the UN phase shares common features with the spin-1 polar phase,
the spin-2 BN phase
represents a more drastic departure from the familiar defect families.

Mathematically, topologically distinguishable
vortices correspond to the conjugacy classes of the first homotopy group
$\pi_1(\M)$ of the order-parameter space $\M$, and are determined by
the group of transformations that keep the order parameter
invariant~\cite{mermin_rmp_1979}. Any BN spinor can be found by applying a
condensate phase $\tau\in\U(1)$ and a spin rotation $\R\in\SO(3)$ to
$\zeta^\mathrm{BN}$.  The fourfold discrete symmetry means
that $\zeta^\mathrm{BN}$ is left invariant by transformations
in the eight-element group
$\Dt_4$ that combines the dihedral-4 subgroup of $\SO(3)$ with elements of
$\U(1)$ and factorizes $\U(1)\times\SO(3)$ to form $\M$.
By lifting $\SO(3)$ to $\SU(2)$ to form a simply connected covering
group, the conjugacy classes of
$\pi_1(\M)$ are obtained using standard techniques~\cite{supplemental}:
$\{(n,\mathbf{1})\}$,
$\{(n,-\mathbf{1})\}$,
$\{(n,\pm i\sx), (n,\pm i\sy), (n, \pm i\sz)\}$,
$\{(n+1/2,\sigma), (n+1/2,-i\sz\sigma)\}$,
$\{(n+1/2,-\sigma), (n+1/2,i\sz\sigma)\}$, and
$\{(n+1/2,\pm i\sx\sigma),(n+1/2,\pm i\sy\sigma)\}$,
where the Pauli matrices $\sigma_{x,y,z}$ and
$\sigma\equiv\left(\mathbf{1}+i\sigma_z\right)/\sqrt{2}$ represent the
$\SU(2)$ part of the $\pi_1(\M)$ elements, and $n$ in the $\U(1)$ part
is an integer.
For $n=0$, we then identify: $(i)$~the vortex-free state,
$(ii)$~integer spin vortex, $(iii)$~spin HQV,
$(iv)$~HQV with $\pi/2$ spin rotation,
$(v)$~HQV with $3\pi/2$ spin rotation, and
$(vi)$~HQV with $\pi$ spin rotation.

We now show that a rich phenomenology of core states, resulting
from the proliferation of spin-2 phases, appears in the BN defects.
We consider the $(1/2,\sigma)$ vortex [case $(iv)$],
which is the simplest BN vortex that carries a mass circulation, and fully
characterize its interaction-dependent core structures.
A prototype wave function for the vortex can be constructed by
applying the corresponding
condensate-phase winding and spin rotation to $\zBN$:
$\zeta^{1/2,\sigma} = e^{i\phi/2}e^{-iF_z\phi/4}\zBN
= (1,0,0,0,e^{i\phi})^T/\sqrt{2}$, where $\phi$ is the azimuthal coordinate.
The energetically stable core structure is determined by numerically
minimizing the energy in the frame rotating with frequency $\Omega$,
corresponding to rotation of the system, e.g., from a stirring potential.
This is done by propagating the Gross-Pitaevskii equations~\cite{javanainen_jpa_2006} derived from
Eq.~\eqref{eq:hamiltonian} in imaginary time, taking $\zeta^{1/2,\sigma}$
as the initial state, including a global spin rotation to ensure
nonzero population in all spinor components.
We take the system-rotation axis to coincide with the
effective magnetic field, but will later relax this assumption.
We choose $Nc_0 = 5000\hbar\omega\ell^3$
(where $\ell=\sqrt{\hbar/m\omega}$), keep $q<0$ fixed, and vary $c_2$
and $c_4$.
For $^{87}$Rb, measurements for $c_0$~\cite{klausen_pra_2001}, $c_2$
and $c_4$~\cite{widera_njp_2006} predict a polar ground state with
$c_2/c_0 \simeq 0.0103$ and $c_4/c_0 \simeq -0.0055$.
Measurement uncertainties, however, cross over
into the cyclic regime.
Also $^{23}$Na is predicted to exhibit polar
interactions~\cite{kawaguchi_physrep_2012}.

We find four different energetically
stable core structures for the $(1/2,\sigma)$ vortex, which
depend on the interaction parameters $c_{0,2,4}$ as indicated in
Fig.~\ref{fig:cores}.
When the interaction strengths are comparable in magnitude, the
superfluid density is depleted on the vortex line singularity [the ($o$)
core]. However, usually $|c_{2,4}|<c_0$ and the core may
remain superfluid. This can be understood from the density, spin, and singlet
healing lengths---$\xi_n=\hbar/\sqrt{2m|c_0|n}$,
$\xi_F=\hbar/\sqrt{2m|c_2|n}$,
and $\xi_A=\hbar/\sqrt{2m|c_4|n}$---which determine the size of
defect cores that take the wave function out of the ground state of a
uniform superfluid.
By expanding to the largest healing length,
the core can lower its gradient energy.
When $c_2$ is small, such that $\xi_A\lesssim\xi_F$, the vortex
develops a FM core [$(\alpha)$ in Fig.~\ref{fig:cores}]  as energy relaxes.
As $c_2$ increases, maintaining a
FM core becomes increasingly costly, and it eventually gives way to
a cyclic core $(\beta)$ as $\xi_A\gtrsim\xi_F$.  However, for
sufficiently large $c_2$, a third
superfluid core $(\gamma)$ approaching the UN phase on the singular line
becomes approximately energetically degenerate, and coexists, with the
$(\beta)$ core. The relaxed structure is sensitive to the initial state.

The measured~\cite{klausen_pra_2001,widera_njp_2006} interaction
strengths for $^{87}$Rb predict an $(\alpha)$ core, close to the
$(\alpha)$-$(\beta)$ boundary. For $^{23}$Na, the
interactions~\cite{kawaguchi_physrep_2012} similarly predict a $(\beta)$
core close to the transition to bistability with the $(\gamma)$
core.
The $(\beta)$ and $(\gamma)$ cores are especially intriguing,
exhibiting a complex mixing of phases that breaks the axial symmetry
(Fig.~\ref{fig:cores}).
The cyclic $(\beta)$ core exhibits a triangular cross section
shown in Fig.~\ref{fig:cores}.
The deformation arises due to the mismatch between the rectangular-brick
BN symmetry and the tetrahedral symmetry of the cyclic
phase~\cite{kobayashi_arxiv}.
In Fig.~\ref{fig:cores}, the system-rotation axis is taken to coincide
with the direction of the effective field corresponding to the Zeeman
shift, which fixes the orientation of the order parameter. Tilting the
vortex line relative to the order parameter, we find an
orientation-dependent instability.  As the angle approaches $\pi/2$,
the smooth connection of the BN and cyclic point-group symmetries is
no longer possible, and the $(1/2,\sigma)$ vortex becomes unstable,
giving way to a singly quantized vortex~\cite{supplemental}.

In the $(\gamma)$ core, the condensate approaches the UN phase
on the line singularity, but exhibits a nonzero $\absF$ as a
result of the Zeeman energy.  The core exhibits twofold symmetry, with
cyclic and BN phases appearing offset on either side of the singular
line as indicated by $|A_{30}|$ and $|A_{20}|$ in
Fig.~\ref{fig:cores} (bottom left).
This complex phase mixing is captured by the parametrization
$\zeta^{(\gamma)} =
\left(\sqrt{1-g_0^2-g_{-2}^2},0,g_0(\rho),0,g_{-2}(\rho)e^{i\phi}\right)^T$,
with only three nonempty spinor components.
The functions $g_0$
and $g_{-2}$ satisfy $g_0(\rho\to\infty)=0$, $g_{-2}(\rho\to\infty)=1/\sqrt{2}$,
and $g_{-2}(\rho\to0)=0$. Then $|A_{30}|^2 =
-12\cos(\phi)f(\rho)+h(\rho)$, where $f$ and $h$ are functions of
$g_{0,-2}$, and the cosine term explains the local extrema.
The value of $g_0$ on the vortex line is determined by energy
minimization.

A striking consequence of the discrete BN symmetry is that the
elements of $\pi_1(\M)$, representing topological charges of line defects,
do not all commute.
Colliding vortices with commuting charges may pass through
each other or reconnect without leaving traces of the process.  For
noncommuting vortices, these processes are forbidden by
topology~\cite{poenaru_jphys_1977,mermin_rmp_1979}.
They must instead reconnect by forming a rung vortex connecting the
resulting defects.  Non-Abelian defects have been proposed in
BN liquid crystals~\cite{poenaru_jphys_1977} and
theoretically classified in cyclic spin-2
BECs~\cite{makela_jpa_2003,semenoff_prl_2007,huhtamaki_pra_2009,kobayashi_prl_2009,mawson_pra_2015}
and in certain spin-3 phases~\cite{barnett_pra_2007}.

We propose a controlled method of preparing a non-Abelian vortex pair and
numerically demonstrate how it could realize a non-Abelian vortex
reconnection in a BN spin-2 BEC. A nonoverlapping perpendicular vortex
pair could be imprinted by a two-photon transition via an intermediate
atomic level, with the singular phase profile of the driving
electromagnetic field of each transition representing one
vortex~\cite{ruostekoski_pra_2005}.
In a non-Abelian pair, however, the vortices are, in
general, entangled in a complex way, making it challenging to imprint them,
as the wave function no longer can be expressed as a combination of simple
quantized vortex lines in individual spinor components. This is because
the transformations of the BN order parameter corresponding to the
$\SU(2)$ charges of each vortex line combine nontrivially. We
therefore propose a two-step protocol where the magnetic field is
rotated between the imprinting of the two vortices, making use of
simple expressions of the vortex lines
in different spinor basis representations~\cite{supplemental}. We
consider a $\pi/2$ rotation that allows simple
representations for both $(1/2,\sigma)$ and $(1/2,i\sigma_x\sigma)$ vortices
[corresponding to the different HQV classes $(iv)$ and $(vi)$] that do
not commute. Moreover, imprinting the
orientation of the vortex lines always along the z axis of the
changing spinor basis prepares the vortex lines perpendicular to each
other, providing an ideal starting point for reconnection dynamics.
The method can utilize the existing
techniques of phase imprinting in spinor BECs by Raman
transitions~\cite{hansen_optica_2016}, and can also be applied, e.g.,
in the cyclic phase.

\begin{figure}[tb]
  \centering
  \includegraphics[width=\columnwidth]{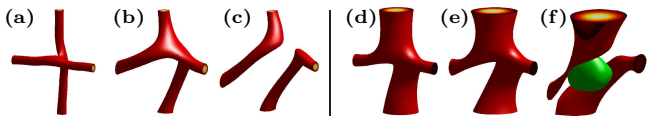}
\caption{Reconnection of (a)--(c) two $(1/2,\sigma)$ vortices,
  and (d)--(f) non-Abelian $(1/2,\sigma)$ and $(1/2,i\sigma_x\sigma)$
  vortices. Red $\absF$ isosurfaces indicate FM vortex cores. Green
  $|A_{30}|$ isosurface indicates the appearance of a rung vortex with
  uniaxial nematic core. Here $Nc_0=1000\hbar\omega\ell^3$,
  $c_0/c_2=100$, $c_0/c_4=-10$, and $q=-0.05\hbar\omega$.
  }
\label{fig:reconnection}
\end{figure}

The corresponding reconnection dynamics is simulated
by propagating the Gross-Pitaevskii equations, including a weak
damping~\cite{supplemental}.
Figure~\ref{fig:reconnection} shows the reconnection and emergence of a rung
vortex with UN core, characteristic of non-Abelian dynamics. For
comparison we also show the reconnection of two $(1/2,\sigma)$
vortices, which trivially commute, and reconnect without
forming a rung.

A distinguishing feature of UNs is the existence of
topologically stable point
defects~\cite{mermin_rmp_1979,pismen,stoof_prl_2001,ruostekoski_prl_2003},
which have very recently been observed in spinor BEC
experiments~\cite{ray_science_2015}. Such a defect
corresponds to a radial hedgehog texture of the nematic axis.
In the spin-1 BEC, the point defect is predicted to relax into a
HQV ring---an Alice ring---as a consequence of the
hairy-ball theorem when the defect develops a superfluid
core~\cite{ruostekoski_prl_2003}. Alice rings are known in special
electrodynamics models of
high-energy physics~\cite{schwarz_npb_1982,alford_npb_1991}.
Despite considerable
experimental efforts, the vortex ring has not been observed
in the spin-1 BEC~\cite{ray_nature_2014}, as on experimental time scales
the size of the ring has remained too small to be
detected~\cite{tiurev_pra_2016}.

In the BN condensate, the
discrete order-parameter symmetry prohibits the formation of an isolated
topologically stable point defect~\cite{mermin_rmp_1979,pismen}: any
such defect
would be attached to a singular line. This is similar to the Dirac magnetic
monopole, whose analog in the FM spin-1
BEC~\cite{savage_pra_2003,tiurev_pra_2016}
was recently prepared in experiment~\cite{ray_nature_2014}.

\begin{figure}[tb]
  \centering
  \includegraphics[width=\columnwidth]{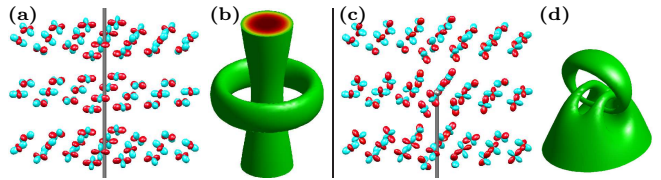}
\caption{(a), (c):  BN order parameter exhibiting a
  radial hedgehog texture [cyan (light gray) lobes align with $\rhat$]
  associated with a singular spin vortex indicated by a vertical line.
  (b), (d):  Isosurface of $|A_{30}|$
  showing the corresponding UN cores of spin vortices resulting from
  energy relaxation. In (c), a more elaborate construction allows the
  vortex to terminate at the point defect~\cite{supplemental}.}
\label{fig:point-defect}
\end{figure}

The simplest way to construct a point defect associated with a vortex
line in the BN spin-2 BEC is to align one of the principal axes of the
order parameter with $\rhat$, as illustrated in
Fig.~\ref{fig:point-defect}(a) using a spherical-harmonics
representation~\cite{supplemental}. This
creates a singular spin-vortex line along the $z$ axis, which develops
a UN core as energy relaxes [Fig.~\ref{fig:point-defect}(b)].
Additionally, energy relaxation causes the point-defect texture in the BN
order parameter to spontaneously deform into a vortex ring, encircling the line
defect.  We identify this as a spin HQV [class $(iii)$].
We simulate the time evolution of the point defect for a spin-2 $^{87}$Rb BEC
and find that it provides a promising system where the
intriguing deformation of a point defect into an Alice ring could be
experimentally observed.
In spin-1 BEC point-defect experiments~\cite{ray_nature_2014} the
size of the ring was estimated to be $\sim
0.2\mu$m~\cite{tiurev_pra_2016}.
In the spin-2 BEC, we numerically find characteristically faster
expansion resulting in an order of magnitude larger ring size than in
spin-1 $^{23}$Na or $^{87}$Rb BEC over experimental time scales, which
is already within an achievable measurement
resolution~\cite{ray_nature_2014}.

In conclusion, we have characterized the coexisting superfluid core
structures of singular defects in a BN spin-2 BEC, and proposed how
vortices exhibiting non-Abelian reconnection dynamics may be prepared
in experiment. The realization of BN defects
in the atomic system opens up several intriguing possibilities:
Nucleation dynamics of non-Abelian defects could be studied in the
spinor-BEC Kibble-Zurek mechanism~\cite{sadler_nature_2006} by cooling
through the BEC transition, or ramping the Zeeman shift into the BN
regime.
The connection of defects across a coherent interface between
phases with different broken symmetry,e.g., mimicking the physics of
string-theoretical branes~\cite{bradley_nphys_2008,sarangi_plb_2002}, could
be studied using techniques similar to those proposed for spin-1
BECs~\cite{borgh_prl_2012,borgh_njp_2014}.
This forms a particularly intricate problem when one phase supports
non-Abelian defects.
The BN spin-2 BEC can
also provide an experimentally simple path to non-Abelian turbulence
scenarios~\cite{mawson_pra_2015}. Preparation of a point defect could
lead to the observation of spontaneous deformation of a spherically
symmetric core to an Alice ring.

The data presented can be found in Ref.~\cite{dataset}.

\begin{acknowledgments}
We acknowledge financial support from the EPSRC. The numerical results
were obtained using the
Iridis 4 high-performance computing facility at the University of
Southampton. We thank T.~J.~Sluckin for discussions.
\end{acknowledgments}

\appendix

\setcounter{equation}{0}
\setcounter{figure}{0}
\renewcommand{\theequation}{S\arabic{equation}}
\renewcommand{\thefigure}{S\arabic{figure}}

\section{Supplemental Material}

\noindent In this Supplemental Material we derive the topologically
distinct line defects in the biaxial nematic spin-2 BEC, and discuss
experimental preparation of a state with orthogonal, noncommuting
vortices. We also provide additional details regarding the
ground-state phase diagram, the orientation-dependent instability of
a singular vortex with cyclic core, and the construction of a point
defect as the termination of a vortex line.

\section*{Classification of vortices in the biaxial nematic phase}

\noindent
Here we formally derive the topologically distinct vortex classes in the
BN phase of the spin-2 BEC, using the homotopy
theory of defects~\cite{mermin_rmp_1979} and following the derivation for
liquid crystals~\cite{poenaru_jphys_1977} and applying the formalism
and methods of Ref.~\cite{semenoff_prl_2007}.
In the atomic spin-2 BEC,
the presence of a condensate phase must also be taken into account,
and leads to vortices with fractionally quantized superfluid circulation.

\begin{figure}[b]
  \centering
  \includegraphics[width=\columnwidth]{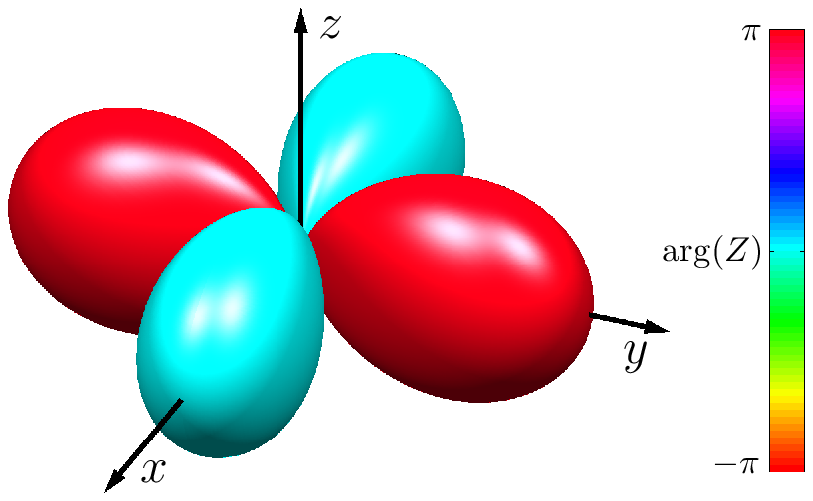}
\caption{Spherical-harmonics representation,
  $Z^\mathrm{BN}=\sum_{m=-2}^{+2}Y_{2,m}\zeta^{\mathrm{BN}}_m$, of the
  representative BN order parameter $\zeta^{\mathrm{BN}}$ in
  Eq.~\eqref{eq:zetaB}. 
  Color represents the complex phase.
}
\label{fig:order-parameter}
\end{figure}

A representative BN spinor is
\begin{equation}
  \label{eq:zetaB}
  \zBN = \frac{1}{\sqrt{2}}\fivevec{1}{0}{0}{0}{1},
\end{equation}
where the spinor components correspond to the spin projection onto the
$z$ axis.
The biaxial symmetry can most easily be seen by representing the order
parameter as a linear combination of spherical
harmonics~\cite{kawaguchi_physrep_2012} 
$Z(\theta,\varphi)=\sum_{m=-2}^{+2}Y_{2,m}(\theta,\varphi)\zeta_m$. Inserting
$\zBN$, we find 
$Z^\mathrm{BN}=\sqrt{15/16\pi}\sin^2\!\theta\cos(2\varphi$),
shown in Fig.~\ref{fig:order-parameter}.

For the purposes of classifying topological defects, it is easier
to work in the Cartesian representation, where the spin-2 order
parameter is a symmetric, traceless rank-2 tensor with
components~\cite{semenoff_prl_2007,kawaguchi_physrep_2012}: 
\begin{equation}
  \label{eq:chi-def}
  \begin{split}
    \chi_{xx} = 
       \frac{\zeta_2+\zeta_{-2}}{\sqrt{2}} &- \frac{\zeta_0}{\sqrt{3}},\\
    \chi_{yy} = 
       -\frac{\zeta_2+\zeta_{-2}}{\sqrt{2}} &- \frac{\zeta_0}{\sqrt{3}},\\
    \chi_{xy} = i\frac{\zeta_2-\zeta_{-2}}{\sqrt{2}},& \quad
    \chi_{xz} = -\frac{\zeta_1-\zeta_{-1}}{\sqrt{2}},\\
    \chi_{yz} = -i\frac{\zeta_1+\zeta_{-1}}{\sqrt{2}},& \quad
    \chi_{zz} = \frac{2}{\sqrt{3}}\zeta_0,
  \end{split}
\end{equation}
where we have used the normalization 
\begin{equation}
  \label{eq:chi-norm}
  \frac{1}{2}\Tr(\chi^\dagger\chi) = 1.
\end{equation}
For the BN spinor~\eqref{eq:zetaB}, the tensor order parameter takes
the simple form 
\begin{equation}
  \label{eq:chiB}
  \chiB = \threemat{1}{0}{0}
                   {0}{-1}{0}
		   {0}{0}{0}.
\end{equation}
The full set of energetically degenerate BN states can be found by
applying $\SO(3)$ rotations, represented by orthogonal $3\times3$
matrices $\R$, and a 
$\U(1)$ phase $e^{i\tau}$. An arbitrary BN order parameter can then
be written on the form $\chi = e^{i\tau}\R\chiB\R^T$.

The group of transformations formed from all possible combinations of
$\tau$ and $\R$ is  
$\U(1)\times\SO(3)$.  We now need to find those elements that leave
$\chiB$ invariant. It is immediately 
obvious from Fig.~\ref{fig:order-parameter}, and 
easily verified, that
$\chiB$ is invariant under the identity transformation and the
three diagonal 
$\SO(3)$ matrices that correspond to rotations by $\pi$ around the $x$,
$y$, and $z$ axes, respectively:
\begin{equation}
  \begin{split}
    \mathbf{1} = \threemat{1}{0}{0}{0}{1}{0}{0}{0}{1},& \quad
    \I_x = \threemat{1}{0}{0}{0}{-1}{0}{0}{0}{-1}, \\
    \I_y = \threemat{-1}{0}{0}{0}{1}{0}{0}{0}{-1},& \quad
    \I_z = \threemat{-1}{0}{0}{0}{-1}{0}{0}{0}{1}.
  \end{split}
\end{equation}
The element
\begin{equation}
    \C = \threemat{0}{1}{0}{-1}{0}{0}{0}{0}{1}
\end{equation}
in $\SO(3)$ represents the $\pi/2$ counter-clockwise rotation around
the $z$ axis, which takes $x \to y$ and $y \to -x$ and thus permutes the
nonzero diagonal elements of $\chiB$. The order parameter is
invariant under $\C$ in combination with
a $\pi$ rotation of the $\U(1)$ phase, $\chiB=e^{i\pi}\C\chiB\C^T$. 
This transformation forms the element $\Ct = (e^{i\pi},\C)$ in 
$\U(1) \times \SO(3)$. 
In the same way, we can construct all remaining transformations that
leave $\chiB$ unchanged as products of $\I_{x,y,z}$ and $\Ct$
(corresponding to $\pi$ rotations around $(\xhat\pm\yhat)/\sqrt{2}$
and the clockwise rotation by $\pi/2$ around $\zhat$). 
We conclude that $\chiB$ is invariant under the eight-element group
\begin{equation}
  \label{eq:D4t}
  \Dt_4 = \{\mathbf{1},\I_x,\I_y,\I_z,\Ct,\I_x\Ct,\I_y\Ct,\I_z\Ct\},
\end{equation}
formed by the dihedral-4 group $\D_4$ in combination with the element
$\exp(i\pi) \in \U(1)$.
Hence the order-parameter space for the BN phase is
\begin{equation}
  \M = \frac{\U(1)\times\SO(3)}{\Dt_4}.
\end{equation}

To classify the vortices in the BN phase, we must now find 
the first homotopy group $\pi_1(\M)$.  
To this end, we find a simply connected covering group by 
lifting $\SO(3)$ to $\SU(2)$, using the quaternion representation, where
the elements of $\SU(2)$ are $2\times2$ matrices $\bm{\mathsf{Q}} =
e_0\mathbf{1}+i\mathbf{e}\cdot\boldsymbol{\sigma}$, given by four Euler
parameters $(e_0,\mathbf{e})$ and the Pauli matrices
$\sigma_{x,y,z}$~\cite{goldstein}.

As we lift $\SO(3)$, the group $\D_4$ lifts into the
16-element subgroup $\D_4^\star$ of $\SU(2)$ according to 
\begin{equation}
  \begin{split}
    \mathbf{1} &\rightarrow \pm\mathbf{1}\\
    \I_{x,y,z} &\rightarrow \pm i \sigma_{x,y,z}\\
    \C &\rightarrow \pm \frac{1}{\sqrt{2}}\left(\mathbf{1}+i\sigma_z\right)
      \equiv \pm \sigma,
  \end{split}
\end{equation}
and combinations of these.
Hence $\Dt_4$ lifts to the 16-element group
\begin{equation}
  \Dt_4^\star = \{\pm\mathbf{1},\pm i\sx,\pm i\sy,
                  \pm i\sz,\pm\tilde\sigma,
		  \pm i\sx\tilde\sigma, \pm i\sy\tilde\sigma,
		  \pm i\sz\tilde\sigma\},
\end{equation}
where $\tilde\sigma$ includes the $\exp(i\pi)$ phase in analogy with
the definition of $\Ct$.  The full subgroup of
$\U(1)\times\SU(3)$ that leaves the order parameter invariant is thus 
\begin{equation}
  \begin{split}
    H = 
      \{&(n,\pm\mathbf{1}),(n,\pm i\sx),
      (n,\pm i\sy),(n,\pm i\sz),\\
      &(n+1/2,\pm\sigma),
      (n+1/2,\pm i\sigma_{x,y,z}\sigma)\},
  \end{split}
\end{equation}
where each element is characterized by an integer $n$, corresponding
to a $2\pi n$ transformation of the $\U(1)$ part, and an element
of $\Dt_4^\star$, giving 16 elements for each $n$.  Fractional $\U(1)$
winding appears when $n$ combines with the $\U(1)$ part of
$\tilde\sigma$. The group
composition law is given by $(x,f)(y,g) = (x+y,fg)$.  Note that the
group is non-Abelian, since in general $fg \neq gf$.
It is also a discrete group, and it follows
immediately~\cite{mermin_rmp_1979} that $\pi_1(\M)=H$.

Topologically distinct vortices correspond to the
\emph{conjugacy classes} of $\pi_1(\M)$, which, since $H$ is
non-Abelian, may contain more than a single element.  
The conjugacy classes are determined by the $\SU(2)$ part of the
elements of $H$. For each $n\in\mathbb{Z}$ we can then directly
calculate six conjugacy classes of $\pi_1(\M)$:
\begin{equation}
  \label{eq:conjugacy-classes}
  \begin{split}
    (i) \quad &\{(n,\mathbf{1})\} \\
    (ii) \quad &\{(n,-\mathbf{1})\} \\
    (iii) \quad &\{(n,\pm i\sx), (n,\pm i\sy), (n, \pm i\sz)\} \\
    (iv) \quad &\{(n+1/2,\sigma), (n+1/2,-i\sz\sigma)\} \\
    (v) \quad &\{(n+1/2,-\sigma), (n+1/2,i\sz\sigma)\} \\
    (vi) \quad &\{(n+1/2,\pm i\sx\sigma),(n+1/2,\pm i\sy\sigma)\}
  \end{split}
\end{equation}
Recalling the rotations of the BN order parameter corresponding to each
element of $\Dt_4^\star$, we then identify the following distinct
types of vortices for the case $n=0$: $(i)$ the trivial element, $(ii)$
singly quantized spin vortex,  $(iii)$ half-quantum spin vortex, $(iv)$
half-quantum vortex 
with $\pi/2$ spin rotation, $(v)$ half-quantum vortex with $3\pi/2$ spin
rotation, and $(vi)$ half-quantum vortex with $\pi$ spin rotation. For $n
\neq 0$ we correspondingly get three types of integer vortex,
involving $0$, $2\pi$ and $\pi$ spin rotations, respectively. In
addition we get three types of half-integer vortices with $\pi/2$,
$-\pi/2$ and $\pi$ spin rotations.

\section*{Vortex preparation and time evolution}

Here we consider construction of a wave function representing a pair
of noncommuting vortices, and show how such a state could be prepared in
experiment. Specifically we construct the wave function of coexisiting
$(1/2,\sigma)$ and $(1/2,i\sigma_x\sigma)$ vortices whose reconnection
is shown in the main text.

Vortex lines with $(1/2,\sigma)$ and $(1/2,i\sigma_x\sigma)$ charges
oriented along the $z$ axis can be
separately constructed by applying the corresponding
condensate-phase and spin rotations to Eq.~\eqref{eq:zetaB} as
\begin{equation}
  \label{eq:1/2-1/4}
  \zeta^{1/2,\sigma} =
  e^{i\phi/2}e^{-iF_z\phi/4}\zBN = \frac{1}{\sqrt{2}}(1,0,0,0,e^{i\phi})^T
\end{equation}
and
\begin{equation}
  \label{eq:1/2-1/2}
  \begin{split}
    \zeta^{1/2,i\sigma_x\sigma} 
    &= e^{i\phi/2}e^{-i\frac{F_x+F_y}{\sqrt{2}}\frac{\phi}{2}}\zBN\\
    &= \frac{e^{i\phi/2}}{\sqrt{2}}
       \fivevec{\cos\frac{\phi}{2}}{e^{-i\pi/4}\sin\frac{\phi}{2}}
 	       {0}
	       {-e^{i\pi/4}\sin\frac{\phi}{2}}{\cos\frac{\phi}{2}},
  \end{split}
\end{equation}
respectively, where $\phi$ is the azimuthal angle. Note how in the
latter case, the spin transformation rotates the order parameter about
the $(\xhat+\yhat)/\sqrt{2}$ as represented in
Fig.~\ref{fig:order-parameter}. 

To construct wave function simultaneously representing
both vortices, we start from Eq.~\eqref{eq:1/2-1/4}, containing the
$(1/2,\sigma)$ vortex, and then add the $(1/2,i\sigma_x\sigma)$
vortex. 
In this case, however, the order parameter in which the
$(1/2,i\sigma_x\sigma)$ is to be added is not spatially uniform.
The axis for the
spin rotation corresponding to the $i\sigma_x\sigma$
$\SU(2)$ charge is therefore no longer constant, but depends on the position
relative to the $(1/2,\sigma)$ vortex. 
We denote the azimuthal angle around the
$(1/2,\sigma)$ vortex by $\phi_1$, and choose coordinates such that
the new vortex line is some distance away in the $\phi_1=\pi$
direction (for simplicity we first assume the vortex lines to be
parallel).  Denoting the 
azimuthal angle relative to the added $(1/2,i\sigma_x\sigma)$ by
$\phi_2=0$, we can find the two-vortex wave function as
\begin{equation}
  \label{eq:noncommuting}
  \begin{split}
  \zeta^{\mathrm{pair}} 
    &= e^{i\phi_2/2}e^{-i\nhat(\phi_1)\cdot\mathbf{\hat{F}}\phi_2/2}\zeta^{1/2,\sigma}\\
    &= \frac{i}{2\sqrt{2}}
       \fivevec{(1+e^{i\phi_2})}{(1-e^{i\phi_2})e^{i\phi_1/4}}{0}
               {-(1-e^{i\phi_2})e^{3i\phi_1/4}}{-(1+e^{i\phi_2})e^{i\phi_1}}.
  \end{split}
\end{equation}
Here $\mathbf{\hat{F}}$ is the vector of spin-2 Pauli matrices. For
parallel vortex lines, the solution is exact.  However, we are
ultimately 
interested in nonparallel vortex lines that exhibit reconnection.
Rotating the direction of the  $(1/2,i\sigma_x\sigma)$ vortex line
(correspondingly rotating the axis defining $\phi_2$),
Eq.~\eqref{eq:noncommuting} becomes approximate, corresponding to a
rapidly relaxing excitation.  Equation~\eqref{eq:noncommuting} for
perpendicular vortex lines forms the initial state for the non-Abelian
reconnection shown in Fig.~2 of the main text.

Vortex lines in the individual components of a spinor BEC can be
prepared using Raman transitions~\cite{hansen_optica_2016}, where the 
singular phase profile of the Raman laser is transferred to the
condensate.  However, Eq.~\eqref{eq:noncommuting} is not 
expressed in terms of simple quantized vortex lines in the spinor
components and is therefore not directly amenable to imprinting using
the Raman process.  
To find an imprinting scheme, we first note as a general
property of the point-group symmetry that the order-parameter
rotations that make up a vortex have simplified representations when
the spinor is expressed in the basis of spin quantization along the
axis of spin rotation.  Accordingly applying the corresponding
spinor-basis transformation 
to Eq.~\eqref{eq:1/2-1/2}, the $(1/2,i\sigma_x\sigma)$ vortex is
expressed as
\begin{equation}
  \begin{split}
    \tilde{\zeta}^{1/2,i\sigma_x\sigma} &= 
    e^{-i\frac{F_x-F_y}{\sqrt{2}}\frac{\pi}{2}}\zeta^{1/2,i\sigma_x\sigma} \\
    &= \frac{1}{\sqrt{2}}
    \fivevec{0}{e^{-3i\pi/4}}{0}{e^{-i\pi/4}e^{i\phi}}{0}.
  \end{split}
\end{equation}
Here, we have introduced $\tilde{\zeta}$ for 
the spinor represented in the rotated basis.
Similarly transforming Eq.~\eqref{eq:noncommuting}, we
find that close to $\phi_1=\pi$ it is well approximated by
\begin{equation}
  \left.\tilde{\zeta}^{\mathrm{pair}}\right|_{\phi_1\simeq\pi} \simeq \frac{1}{\sqrt{2}}
               \fivevec{0}{e^{-i\pi/4}}{0}{e^{i\pi/4}e^{i\phi_2}}{0},
\end{equation}
exhibiting a singly quantized vortex line only in
$\tilde{\zeta}_{-1}$, corresponding to the $(1/2,i\sigma_x\sigma)$
vortex. 
This
suggests a two-step imprinting scheme where first the $(1/2,\sigma)$
vortex is prepared according to Eq.~\eqref{eq:1/2-1/4}. The effective
magnetic field, corresponding to the Zeeman shift, is rotated by
$\pi/2$, changing the spinor basis. The $(1/2,i\sigma_x\sigma)$ vortex
is then added by additionally preparing a vortex line in
$\tilde{\zeta}_{-1}$. This can proceed via population transfer to an
intermediate level and then back, using the appropriate laser
configuration to imprint the vortex line.
In the basis corresponding to the final direction of the
magnetic field, the two-vortex wave function is then approximated by
the imprinted state 
\begin{equation}
  \label{eq:imprinted}
  \tilde{\zeta}^{\mathrm{pair}} \approx \tilde{\zeta}^{\mathrm{imp}}
  = \frac{1}{4\sqrt{2}} 
    \fivevec{(1+e^{i\phi_1})}
            {2i(-1+e^{i\phi_1})}
            {-\sqrt{6}(1+e^{i\phi_1})}
            {-2ie^{i\phi_2}(-1+e^{i\phi_1})}
            {(1+e^{i\phi_1})}.
\end{equation}
In regions away from $\phi_1\simeq\pi$ this wave function leaves the BN phase,
corresponding to an excitation that relaxes under dissipation.  In the
two-step protocol, each vortex can be prepared using the Raman
process. By imprinting vortex lines along the $z$ axis of the changing
spinor basis, perpendicular vortex lines that exhibit non-Abelian
reconnection may be prepared.

In order to show the non-Abelian reconnection of vortices with
noncommuting charges, we follow the time evolution of the state with
two orthogonal vortices described by Eq.~\eqref{eq:noncommuting}. We 
numerically propagate the coupled Gross-Pitaevskii equations derived
from the Hamiltonian density using a split-operator method.  In doing
so, we include a weak phenomenological dissipation by taking the time $t \to
(1-i\eta)t$~\cite{gardiner_jphysb_2002}.  Here we choose $\eta=10^{-3}$.

\section*{Interactions and steady-state solutions in the spin-2 BEC}

Here we give details of the interaction terms in the Gross-Pitaevskii
mean-field Hamiltonian and 
provide explicit definitions of the steady-state solutions that appear as
ground states in
addition to the BN in the phase diagram [Fig.~1(a) of the main text] for
quadratic Zeeman shift $q<0$. A full derivation of all steady-state
solutions of the uniform spin-2 BEC (including those that do not form
the ground state for any parameter range) can be found in
Ref.~\cite{kawaguchi_physrep_2012}.

The Hamiltonian density is given by Eq.~(1) in the main text.  The
strengths $c_{0,2,4}$ of the three nonlinear interaction terms are
derived from the $s$-wave scattering lengths $a_{0,2,4}$ that
correspond to the three scattering channels of colliding spin-2 atoms, as $c_0 =
4\pi\hbar^2(3a_4+4a_2)/7m$, $c_2 = 4\pi\hbar^2(a_4-a_2)/7m$, and 
$c_4 = 4\pi\hbar^2(3a_4-10a_2+7a_0)/7m$, where $m$ is the atomic
mass.  For $q<0$, the BN state forms the ground state of the uniform
system when  $c_2 > c_4/20$ and $c_4 < 10|q|/n$, where $n$ is the
atomic density.  This is 
the case for the most commonly used atoms for spinor-BEC experiments:
Measurements in the spin-2 manifold of $^{87}$Rb give
$a_2-a_0=(3.51\pm0.54)a_B$ and $a_4-a_2=(6.95\pm0.35)a_B$ (where $a_B$
is the Bohr radius)~\cite{widera_njp_2006}, such that
$c_4/c_2\simeq-0.54$. Also the spin-2 manifold of $^{23}$Na is
predicted to exhibit the same ground state with $a_0=34.9\pm1.0$,
$a_2=45.8\pm1.1$, and $a_4=64.5\pm1.3$~\cite{kawaguchi_physrep_2012}, giving
$c_4/c_2\simeq-1.1$. 

We now briefly define the remaining steady-state solutions and detail
when they form the ground state in the
$q<0$ phase diagram. The spin-2 FM state is the ground state when $c_2
< c_4/20$ for $c_4 < 10|q|/n$, and when $c_2 < |q|/(2n)$ otherwise. The 
spinor wave function is then $\zeta^{\mathrm{FM}} = (e^{i\chi_{+2}},0,0,0,0)^T$
[or $\zeta^{\mathrm{FM}} = (0,0,0,0,e^{i\chi_{-2}})^T$, which has the
same energy for $p=0$]. In this state, 
$\absF=2$, and the energy per particle is $\epsilon = c_0n/2 +
2c_2n + 4q$.

When $c_4n \geq 10|q|$ and $c_2 > |q|/(2n)$, two additional
steady-state solutions appear~\cite{kawaguchi_physrep_2012}.  The first is  
\begin{equation}
  \zeta^{\mathrm{C}} =
  \left(e^{i\chi_{+2}}\sqrt{\frac{1+f_z}{3}},0,0,e^{i\chi_{-1}}\sqrt{\frac{2-f_z}{3}},0\right)^T
\end{equation}
[or $\zeta^{\mathrm{C}} =
(0,e^{i\chi_{+1}}\sqrt{(2+f_z)/3},0,0,e^{i\chi_{-2}}\sqrt{(1-f_z)/3})^T$],
with $\absF = |f_z| = |q/(c_2n)|$ and energy 
$\epsilon = c_0n/2 + 2q - q^2/(2c_2n)$.  
The other is 
\begin{widetext}
\begin{equation}
  \zeta^{\mathrm{C^\prime}} =
  \left(ie^{i\chi}\frac{\sqrt{1-10q/(c_4n)}}{2},
    0,\sqrt{\frac{1+10q/(c_4n)}{2}},0,
    ie^{-i\chi}\frac{\sqrt{1-10q/(c_4n)}}{2}\right)^T,
\end{equation}
\end{widetext}
with $\absF=0$ and energy $\epsilon = c_0n/2 + 2q - 10q^2/(c_4n)$. The
$\mathrm{C}$ ($\mathrm{C}^\prime$) steady-state solution is the ground
state when $c_2 < c_4/20$ ($c_2 > c_4/20$). Both spinors continuously approach
(different representations of) the cyclic order parameter as $q \to 0^-$.

\section*{Orientation-dependent instability of 
  $(1/2,\sigma)$ vortex with cyclic core}

\begin{figure}[tb]
  \centering
  \includegraphics[width=\columnwidth]{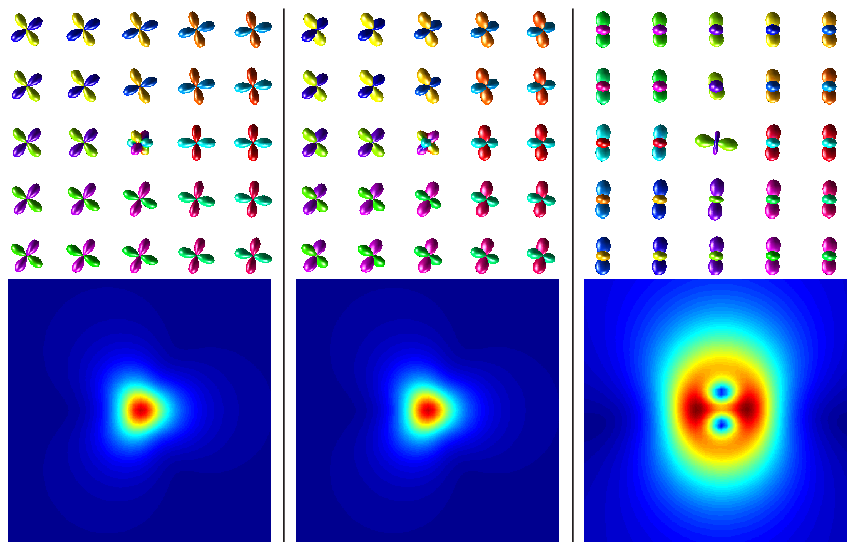}
  \caption{Relaxed state of an initial $(1/2,\sigma)$
    vortex when the quadratic Zeeman shift corresponds to magnetic field
    making an angle $0$ (Left), $\pi/4$ (Middle), and $\pi/2$ (Right)
    with the trap rotation axis. (Top) Spherical harmonics
    representation of the relaxed vortex state. (Bottom) $|A_{30}|$
    showing the structure and shape of the core. The triangular shape of
    the cyclic $(\beta)$ core remains unchanged until a very large angle
    renders the $(1/2,\sigma)$ vortex unstable, in favor of a singly
    quantized vortex with a core exhibiting mixing of cyclic, UN and BN
    phases.  $Nc_0 = 5000\hbar\omega\ell^3$, $Nc_2=-Nc_4=1000\hbar\omega\ell^3$,
    $q=-0.05\hbar\omega$, and $\Omega = 0.17\omega$. 
  }
  \label{fig:angle}
\end{figure}

Here we study the orientation-dependent instability of the
$(1/2,\sigma)$ vortex with the cyclic $(\beta)$ core that follows from
the breakdown of the smooth connection of the different point-group
symmetries of the vortex core and the bulk superfluid.
The cyclic core breaks axial symmetry as a result of the
incommensurate spin symmetries~\cite{kobayashi_arxiv}, so that its
spatial symmetry 
reflects the threefold discrete spin symmetry of the cyclic
order parameter.  The left panels of Fig.~\ref{fig:angle} show the
order parameter in the 
spherical-harmonics representation along with $|A_{30}|$ for the
stable vortex with the cyclic $(\beta)$ core.  

The orientation of the BN and 
cyclic order parameters is fixed by the quadratic Zeeman shift, 
which corresponds to a magnetic field along the $z$ axis. We have
taken this to coincide with the trap rotation axis, but this need not
in general be the case.  In the middle panel of Fig.~\ref{fig:angle},
the rotation axis has been tilted with respect to the effective
field.  The vortex-core shape remains robust through a wide range of
tilt angles, consistent with the topological origin of the
deformation.  However, when the tilt angle approaches $\pi/2$, the
system can no longer form the continuous connection of the two
point-group symmetries, resulting in a destabilization of the vortex
structure. The $(1/2,\sigma)$ vortex then
gives way to a singly quantized $(1,0)$ vortex shown in the right-hand
panels of Fig.~\ref{fig:angle}.   
The core of the vortex exhibits a highly complex mixing of cyclic, UN,
and BN phases, indicated by $|A_{30}|$ in Fig.~\ref{fig:angle}.  At
the center of the region, the UN phase appears.  This is surrounded by
two cyclic and to BN regions, corresponding to the local maxima and
minima, respectively.

\section*{Construction of point defect as the termination of a vortex line}

In the main text we presented the simplest way of constructing a
point-defect texture of the BN order parameter, resulting in an
associated spin-vortex line along the $z$ axis.  In this construction,
one of the principal axes of the BN order parameter 
(cf.\ Fig.~\ref{fig:order-parameter}) is aligned with the radius vector.
Through a more elaborate construction, we can
avoid the line defect on the positive $z$ axis, illustrated in
Fig.~3(c) of the main text.  The point defect is
still formed by aligning a principal axis with the radius vector.
However, by including a local spin rotation about $\rhat$, the BN
order parameter can remain nonsingular in the entire upper hemisphere.
The point defect then forms the termination point of a spin-vortex
line.  Energy relaxation in this case leads to the complicated structure
of interlocking spin vortex rings shown in Fig.~3(d).

\end{document}